# X-ray photon detection using superconducting resonators in thermal quasi-equilibrium

O. Quaranta[1,a], T. W. Cecil[1], L. Gades[1], B. Mazin[2], and A. Miceli[1,b]

[1] *X-ray Science Division, Argonne National Laboratory, Argonne, Illinois 60439, USA*

[2] *Department of Physics, University of California, Santa Barbara, California 93106, USA*

(Dated: June 4, 2013)

## Abstract

Superconducting resonators have to date been used for photon detection in a non-equilibrium manner. In this paper, we demonstrate that such devices can also be used in a thermal quasi-equilibrium manner to detect X-ray photons. We have used a resonator to measure the temperature rise induced by an X-ray photon absorbed in normal metal and superconducting absorbers on continuous and perforated silicon nitride membranes. We observed two distinct pulses with vastly different decay times. We attribute the shorter pulses to non-equilibrium quasiparticle relaxation and the longer pulses to a thermal relaxation process. In addition, we have measured the temperature dependence of the X-ray induced temperature rise and decay times. Finally, we have measured the resonator sensitivity and energy resolution. Superconducting resonators used in a thermal quasi-equilibrium manner have the potential to be used for X-ray microcalorimetry.

---

[a] Electronic address: oquaranta@aps.anl.gov
[b] Electronic address: amiceli@aps.anl.gov

A photon absorbed in a superconductor can break Cooper Pairs, generating quasiparticles with energy much greater than the superconducting energy gap ($\Delta$). The quasiparticles relax to an equilibrium state by scattering with other quasiparticles or phonons and by recombining into Cooper Pairs [1]. The photon absorption process generates a transient, excess quasiparticle population above the thermal equilibrium population, which will decay on a time scale characteristic of the material and the bath temperature [2]. This non-equilibrium population of quasiparticles changes the complex conductivity of the superconductor, ($\sigma_{1,2}$), and can be sensed by patterning a film in the form of a resonator and monitoring the time evolution of both the resonance frequency ($f_r$) and the internal quality factor ($Q_i$) [3,4]. Non-equilibrium photon detectors based on this principle are known as Microwave Kinetic Inductance Detectors (MKIDs). However, the response of a resonator is nearly identical to either an excessive non-equilibrium quasiparticle population from photon absorption or an increase in equilibrium thermal quasiparticle population due to a temperature change [5,6]. This equivalence is routinely used to statically measure the response and uniformity of large arrays of resonators as a function of temperature. However, as we demonstrate here, this equivalency enables the use a superconducting resonator to detect X-ray photons in a thermal quasi-equilibrium manner.

We design our devices using a superconducting resonator in the lumped element form with the inductive meander shaped to encircle an absorber on a SiN membrane. The thermal circuit diagram is shown in Fig. 1a. In contrast to non-equilibrium X-ray MKIDs [7], the resonator and absorber are not in direct electrical contact. Energy flow between the resonator and absorber is mediated by the phonons in the SiN membrane. The lumped element resonator design allows us to separate the inductor (i.e., the active part of the resonator) from the capacitor. While the inductive meander is shaped to encircle the absorber on the SiN membrane, the interdigitated capacitor (IDC) is off the membrane on the solid wafer, capacitively coupled to the feedline. We have fabricated several configurations of the absorber and the SiN membrane. In all of the devices, the resonator is a single layer of 100 nm thick tungsten silicide (WSi$_2$, $T_c \approx 1.3$ K) on a 0.5 μm SiN membrane. One device has a 300 × 300 × 0.5 μm$^3$ tantalum absorber with a 10 nm thick niobium seed layer. This device lies on a SiN membrane perforated to form an island (Fig. 1b). The other two devices have 0.1 μm thick gold absorber on continuous SiN. On the same chip, gold devices consist of two areas: 300 × 300 μm$^2$ (Gold-300), 200 × 200 μm$^2$ (Gold-200), in addition to devices with no absorber present. The WSi$_2$ is deposited by DC magnetron sputtering from a stoichiometric target and the resonator is patterned with a dry etch. The absorbers are patterned via liftoff. A KOH wet etch is used to etch through the 300 μm thick high-resistivity (>5 kΩ cm) Si wafer and release the continuous suspended SiN membrane. The tantalum device has a final dry etch to form the SiN island.

We have illuminated these devices with uncollimated X-ray photons of ≈ 6 keV from Fe-55 at different bath temperatures and studied the phase pulses from the resonators using a standard IQ mixing setup [3,4]. The measured or loaded quality factors ($Q$) of the resonators are: $9 \times 10^3$, $5.7 \times 10^4$, and $2.7 \times 10^4$ for the Gold-300, Gold-200, and tantalum devices respectively. When illuminated with X-ray photons the devices respond with a decrease of the resonance frequency and internal quality factor that can be seen as a pulse in both phase and amplitude directions. The amplitude pulses were typically smaller than the phase pulses. We present an analysis using only the phase response.

We observe two general classes of pulses, both represented in Fig. 2. The blue solid line pulse is similar to the typical pulses seen in other tungsten silicide resonators fabricated on thick silicon and sapphire substrates [8, 9]. This type of response is compatible with the expected non-equilibrium response for these energies for a given Q, kinetic inductance fraction, and volume [8]. We attribute these pulses primarily to absorption of photons in the SiN membrane with subsequent generation of athermal phonons. These athermal phonons are energetic enough to generate a large number of non-equilibrium quasiparticles, which results in a large response [10, 11] and a very short decay time in tungsten silicide. For the tantalum device on the perforated SiN membrane, these pulses have a second decay time, which is likely due to phonon trapping or thermalization in the SiN, given the very weak thermal link to bath. We do not see this second decay time in the athermal pulses from the gold devices on the continuous membrane. The fast portion of the pulse is compatible with the resonator response time of few μs ($\tau_{res} = Q/\pi f_r$, $Q \approx 10^4$ and $f_r \approx 3$ GHz), which indicates that the quasiparticle recombination time for tungsten silicide is probably much shorter. While this represents a disadvantage from the point of view of non-equilibrium resonators, where the quasiparticles will not live long enough to be properly sensed, it represents an advantage for thermal MKIDs, where the fast thermalization of the resonator should ensure the equilibrium of the resonator with the rest of the system (i.e., absorber and membrane) as well as suppress the generation-recombination noise inside the resonator bandwidth [12].

The longer pulse in Fig. 2 has never been seen in tungsten silicide resonators. The most significant difference is the much longer decay time, more than an order of magnitude longer than previously seen. In the gold devices on a continuous membrane, this type of pulse is much less frequent than the fast pulses: ≈ 10% for Gold-200 and ≈ 15% for Gold-300. Considering the comparable stopping power of gold and SiN at these thicknesses, these percentages are compatible with the ratio of absorber area to membrane area. On the other hand, in the tantalum device with greater X-ray absorption efficiency (~25% at 6 keV) and substantially less SiN, the majority of the pulses are those with the much longer decay times. The gold devices are fit well to a single exponential decay while the tantalum devices are fit with a double exponential. These two exponentials will be discussed later when we examine their dependence with bath temperature.

To understand the nature of these long pulses, the detection mechanism behind a superconducting resonator needs to be considered. A superconductor in thermal equilibrium at a given bath temperature ($T_{bath}$) is characterized by a thermal population of quasiparticles $n_{qp}(T) = 2 N_0 \sqrt{2\pi kT\Delta}\, e^{-\frac{\Delta}{kT}}$, where N₀ is the single spin density of state at the Fermi level and k is Boltzmann's constant. An instantaneous temperature rise (*dT*) in the superconductor will result in an increase in the thermal population of quasiparticles $dn_{qp} = n_{qp}(T_{bath} + dT) - n_{qp}(T_{bath})$. This change in the density of quasiparticles will result in a change in the complex conductivity $(d\sigma_{1,2}/dn_{qp})$ and thus a change in phase given by, [4] $d\theta(T) = (2\gamma Q/\sigma_2(T=0))\left(d\sigma_2(T)/dn_{qp}\right) dn_{qp}$, where $\gamma$ is the kinetic inductance fraction. This represents a direct relationship between temperature rise and the phase change. If we assume that the long decay pulses are of a thermal nature, due to the absorption of the photon in the gold or the tantalum absorber, we can use this relationship to estimate the induced temperature rise in the resonator from the observed phase change. This assumes the knowledge of several material parameters, including $N_0$ and $\Delta$, which are not well known for WSi$_2$. The same result can be obtained by mapping the phase change induced in the resonator by the change of the bath temperature. Starting from a given bath temperature (i.e. a given resonance frequency), we slowly raised the bath temperature by regulating the cryostat in 1 mK steps and measured the corresponding phase change. This creates a calibration curve for the resonator acting as a thermometer. Calibration curves for a tantalum device taken at three different bath temperatures are shown in Figure 3. Using a matched filter, we estimated the pulse height of the long decay pulses, and with the calibration curve, we can calculate the induced temperature rise at different $T_{bath}$ for each device [13]. The resulting temperature rises are shown in Figure 4a.

As expected, the lower heat capacity of the superconducting tantalum absorber produces a much larger temperature rise than that of the normal metal gold absorber. The expected temperature rise in the gold devices given by literature values of gold and SiN heat capacities is within a factor two of the measured temperature rise, which is reasonable given the uncertainties in the heat capacities of our SiN and gold. The expected temperature rise from the tantalum device is a factor of ten smaller than expected assuming the classical Debye phonon heat capacity for tantalum and typical heat capacities for SiN [14]. There are a number of possible reasons for this discrepancy. Since tantalum has a superconducting energy gap, some fraction of the X-ray energy can be trapped in a quasiparticle-phonon bottleneck. Perinati *et al.*, [15] observed that a large fraction of the energy thermalizes on a very long time scale (several milliseconds), and thus would not raise the temperature of the island. There has also been evidence for anomalous heat capacities in tantalum [16]. Finally, the measurements were performed in an ADR cryostat, and it is possible that trapped magnetic flux could result in normal-state zones, which would increase the tantalum heat capacity despite a magnetic shield surrounding the sample box. At the same time, we do not have a good estimate of our SiN membrane heat capacity and thermal conductivity.

The temperature dependence of resonator relaxation or decay time can provide insights into the nature of the relaxation process. The relaxation time of non-equilibrium quasiparticles as a function of temperature is described by [2]:

$$\tau_{rec} = \frac{\tau_0}{\sqrt{\pi}} \left(\frac{kT}{2\Delta}\right)^{5/2} \sqrt{\frac{T_c}{T}}\ e^{\Delta/kT} \quad (1)$$

where $\tau_0$ is a material dependent parameter related to the electron–phonon coupling strength. The relaxation time of non-equilibrium quasiparticles decreases with increasing temperature (except at the lowest temperatures [17]) and materials with lower $T_c$ typically have longer recombination times (tungsten silicide being a notable exception [8, 9]). While $\tau_{rec}$ can be enhanced by $2\Delta$-phonon reabsorption, which depends on the phonon pair-breaking and escape times [18], the overall temperature dependence will remain that of Eq. 1.

In Fig. 4b, we show the temperature dependence of the resonator decay times. In order to model the decay time with temperature, detailed knowledge of the heat capacities and thermal conductances is required, but we can use the observed trends to understand the relaxation process. The two gold devices show decay times that decrease with temperature. The order of magnitude of the decay time and the rate of decay with bath temperature are far from what is predicted by Eq. 1. At the same time, the tantalum device has a slow and fast decay time that shows an overall increase with temperature. In all the devices, the resonator is identical. The difference between the devices is the absorber and the SiN membrane. The increasing decay time in the tantalum devices is a clear departure from Eq. 1. The temperature dependence of the decay times appears to be dependent on the nature of the absorbers (normal versus superconducting) that have heat capacities with different temperature dependences and also on the type of SiN membrane (continuous versus perforated), which have thermal conductance with different temperature dependences as well. The two decays in the tantalum device have similar temperature dependences and are both likely to be of a thermal nature given that this complex electro-thermal system will have multiple time constants. While long athermal tails have been observed in tantalum [19], these tend to be much longer than our detector time constants and may be due to the epoxy used for attachment [20]. The exact nature of these two decay times will require further investigation.

The temperature dependence of the decays implies that in this device we are not generating an excess of non-equilibrium quasiparticles by the absorption of photons directly in the resonator or the substrate [21] or by the injection of quasiparticles in the resonator from a separate absorber in electrical contact [7]. Rather, the resonator is responding to the thermal relaxation of the absorbers. This is in contrast to the work of [22], where they observed an enhancement in the recombination time of an aluminum coplanar waveguide resonator fabricated on a SiN membrane when illuminated directly with optical photons. In our case, we observe two distinct types of pulses. While it is possible that there is an enhancement of the recombination time in the tungsten silicide resonator itself, this is probably still shorter than the resonator response time. The thermal relaxation in these devices seems to depend only on the heat capacity and the thermal conductance of the system. In this letter, the silicon nitride membrane has provided a very weak thermal link to the bath, and it has enabled us to clearly see the thermal response of the superconducting resonators. However, this thermal behavior can also affect the response of resonators on solid substrates. This was pointed out in N. Vercruyssen *et al.*, [22] where they saw an enhancement in resonator decay time for aluminum resonators ($T_c$ = 1.1 K) on a solid SiN/Si substrate in the temperature range of 0.3 to 0.4 K, which is not explained by quasiparticle recombination [2]. Likewise, de Visser *et al.*, [17] saw a similar enhancement of the resonator decay time in a similar temperature range for aluminum resonators on sapphire substrates illuminated with optical photons. Likely, this can be explained by considering the thermal behavior of the system.

Using resonators for X-ray microcalorimetry is potentially attractive given their straightforward multiplexing capabilities. Ultimately, microcalorimeters are limited by thermodynamic fluctuations [23] across the thermal weak links. In our devices, the absorber and resonator are not in direct contact, which will add an additional contribution to the thermodynamic fluctuations. However, it is conceivable to place the absorber and resonator in direct contact. In addition, superconducting resonators have excess phase noise which can be mitigated by better choice of substrate (e.g., silicon membrane using silicon-on-insulator wafers [24, 25]) and design considerations [26, 27]. Further analysis will be required to understand the fundamental limits [28]. However, with our current device we can make an initial examination of the detector performance. Recently Lindeman *et al.*, [29] have proposed a logarithmic sensitivity for an inductive thermometer:

$$\alpha_\theta = Q \frac{T}{L}\frac{dL}{dT} = \frac{T}{2}\frac{d\theta}{dT} \quad (2)$$

where *L* is the total inductance. We can use the curves in Figure 3 to determine the sensitivity as a function of bath temperature for the tantalum device. At $T_{bath}$ = 175 mK, the sensitivity is $\alpha_\theta \approx$ 5. While this sensitivity is smaller than that of a transition-edge sensor, it is comparable to semiconductor thermistors [24]. In addition, the sensitivity can be increased using higher *Q* resonators and operating at a bath temperature closer to $T_c$. Using a matched filter we created a histogram of pulse heights after applying a linear rise time-pulse height correction. The FWHM energy resolution of the Mn Kα line is 127 eV at 175 mK, while the expected resolution [23] given the pulse template and noise power spectrum is 32 eV. The capacitor for this device is on SiN, a source of two-level system that cause increased phase noise, and we anticipate a reduction in noise and baseline resolution by removing the SiN under the capacitor. The FWHM resolution is likely limited by position-dependent response in the tantalum absorber, a well-known issue with absorbers where the thermalization time is not negligible [30]. In particular, the thermalization of superconductors, such as tantalum, is particularly slow and incomplete [15]. However, superconducting tin [31] and semiconducting HgTe [32] have successfully been used as X-ray absorbers using TES and semiconductor thermistors, respectively. Moreover, if the distribution of the gradient of temperature from the absorber to the bath in the silicon nitride is not isotropic, the intrinsic current distribution in the meander portion of the resonator can degrade the resolution. We expect to reduce our baseline noise level (i.e., phase noise) in future designs by removing the layer of SiN underneath the IDC and are pursuing design solutions to address systematic issues.

In conclusion, we have fabricated and characterized a superconducting resonator with X-ray photons that bases its detection mechanism on the measurement of temperature variation induced by photon absorption in gold and tantalum absorbers. In this approach, the quasiparticle population in the resonator remains in thermal quasi-equilibrium with the SiN membrane, as opposed to the traditional MKID detection mechanism where an excess of non-equilibrium quasiparticles in the resonator is measured. We have increased the decay time of the pulses from tungsten silicide resonators by more than an order of magnitude. The superconducting resonators used in a thermal quasi-equilibrium manner have the potential to be used for X-ray microcalorimetry.

The authors would like to thank Vlad Yefremenko, Ralu Divan, Suzanne Miller, Chian Liu for valuable discussions on device fabrication. We thank Yejun Feng for valuable discussions. Use of the Center for Nanoscale Materials was supported by the U. S. Department of Energy, Office of Science, Office of Basic Energy Sciences, under Contract No. DE-AC02-06CH11357. Work at Argonne National Laboratory was supported by the U.S. Department of Energy, Office of Science, Office of Basic Energy Sciences, under Contract No. DE-AC02-06CH11357.

**References**

[1] A. G. Kozorezov, J. K. Wigmore, D. Martin, P. Verhoeve, and A. Peacock, Phys. Rev. B 75, 094513 (2007).

[2] S. B. Kaplan, C. C. Chi, D. N. Langenberg, J. J. Chang, S. Jafarey, and D. J. Scalapino, Phys. Rev. B 14, 4854 (1976).

[3] P. K. Day, H. G. LeDuc, B. A. Mazin, A. Vayonakis, and J. Zmuidzinas, Nature 425, 817 (2003)

[4] B. A. Mazin, Ph.D. Dissertation, California Institute of Technology (2004).

[5] J. Gao, J. Zmuidzinas, A. Vayonakis, P. Day, B. Mazin, and H. Leduc, J Low Temp. Phys. 151, 557 (2008)

[6] J. Gao, Ph.D. dissertation, California Institute of Technology (2008).

[7] B. A. Mazin, B. Bumble, P. K. Day, M. E. Eckart, S. Golwala, J. Zmuidzinas, and F. A. Harrison, Appl. Phys. Lett. 89, 222507 (2006).

[8] T. W. Cecil, A. Miceli, O. Quaranta, C. Liu, D. Rosenmann, S. McHugh, and B. Mazin Appl. Phys. Lett. 101, 032601 (2012).

[9] O. Quaranta, T. W. Cecil, and A. Miceli, IEEE Trans. Appl. Supercond. 23, 2400104 (2013).

[10] D. C. Moore, S. R. Golwala, B. Bumble, B. Cornell, P. K. Day, H. G. LeDuc, and J. Zmuidzinas, Appl. Phys. Lett. 100, 232601 (2012).

[11] L. J. Swenson, A. Cruciani, A. Benoit, M. Roesch, C. S. Yung, A. Bideaud, and A. Monfardini, Appl. Phys. Lett. 96, 263511 (2010).

[12] J. Zmuidzinas, Annu. Rev. Condens. Matter Phys. 3, 169–214 (2012).

[13] We observed a linear correlation in the measured rise time and pulse height in the tantalum device, and we have applied a linear correction. We believe this correlation is a symptom of a position-dependent response because of the slow thermalization time of tantalum.

[14] G. Wang, V. Yefremenko, V. Novosad, A. Datesman, J. Pearson, G. Shustakova, R. Divan, C. Chang, J. McMahon, L. Bleem, A. T. Crites, T. Downes, J. Mehl, S. S. Meyer, and J. E. Carlstrom, Advances in Cryogenic Engineering, Materials, vol. 56, pp. 75–83, (2009).

[15] E. Perinati, M. Barbera, S. Varisco, E. Silver, J. Beeman, and C. Pigot, Rev. Sci. Instrum. 79, 053905 (2008).

[16] G. J. Sellers, A. C. Anderson, and H. K. Birnbaum, Phys. Rev. B 10, 2771-2776 (1974).

[17] P. J. de Visser, J. J. A. Baselmans, P. Diener, S. J. C. Yates, A. Endo, and T. M. Klapwijk, Phys. Rev. Lett. 106, 167004 (2011).

[18] A. Rothwarf and B. N. Taylor, Phys. Rev. Lett. 19, 27 (1967).

[19] M.L. van den Berg, D.T. Chow, A. Loshak, M.F. Cunningham, T.W. Barbee Jr., M. Frank, S.E. Labov, Proc. SPIE 4140, X-Ray and Gamma-Ray Instrumentation for Astronomy XI, 436 (December 13, 2000).

[20] R. Horansky, J. Beall, K. Irwin, and J. Ullom, AIP Conf. Proc. 1185, 733-736 (2009).

[21] B. A. Mazin, B. Bumble, S. R. Meeker, K. O'Brien, S. McHugh, and E. Langman, Optic Express 20, 1503 (2012).

[22] N. Vercruyssen, R. Barends, T. M. Klapwijk, J. T. Muhonen, M. Meschke, and J. P. Pekola, Appl. Phys. Lett. 99, 062509 (2011).

[23] S. H. Moseley, J. C. Mather, and D. McCammon, J. Appl. Phys. 56, 1257 (1984).

[24] R. Kelley, et al., Publ. Astron. Soc. Jpn. 59, S77 (2007).

[25] J. Hubmayr, J. Beall, D. Becker, H.-M. Cho, B. Dober, M. Devlin, A. M. Fox, J. Gao, G. C. Hilton, K. D. Irwin, D. Li, M. D. Niemack, D. P. Pappas, L. Vale, and M. Vissers, Applied Superconductivity, IEEE Transactions on, vol.23, no.3, pp.2400304, June 2013.

[26] J. Gao, M. Daal, A. Vayonakis, S. Kumar, J. Zmuidzinas, B. Sadoulet, B. A. Mazin, P. K.

Day, and H. G. LeDuc, Appl. Phys. Lett. 92, 152505 (2008).

[27] R. Barends, N. Vercruyssen, A. Endo, P. J. de Visser, T. Zijlstra, T. M. Klapwijk, and J. J. A. Baselmans, Appl. Phys. Lett. 97(3), 033507 (2010).

[28] M. Galeazzi and D. McCammon, J. Appl. Phys. 93, 4856 (2003).

[29] M. A. Lindeman, P. Khosropanah, and R. A. Hijmering, J. Appl. Phys. 113, 074502 (2013).

[30] T. Saab, E. Figueroa-Feliciano, N. Iyomoto, S. R. Bandler, J. A. Chervenak, R. L. Kelley, C. A. Kilbourne, F. S. Porter, and J. E. Sadleir, J. Appl. Phys. 102, 104502 (2007).

[31] D. A. Bennett, R. D. Horansky, D. R. Schmidt, A. S. Hoover, R. Winkler, B. K. Alpert, J. A. Beall, W. B. Doriese, J. W. Fowler, C. P. Fitzgerald, G. C. Hilton, K. D. Irwin, V. Kotsubo, J. A. B. Mates, G. C. O'Neil, M. W. Rabin, C. D. Reintsema, F. J. Schima, D. S. Swetz, L. R. Vale, and J. N. Ullom, Rev. Sci. Instrum. 83, 093113 (2012).

[32] P. Dreiske, M. Carmody, C. H. Grein, J. Zhao, R. Bommena, C. A. Kilbourne, R. Kelley, D. McCammon, and D. Brandl, J. Electron. Mater. 39, 1087 (2010).

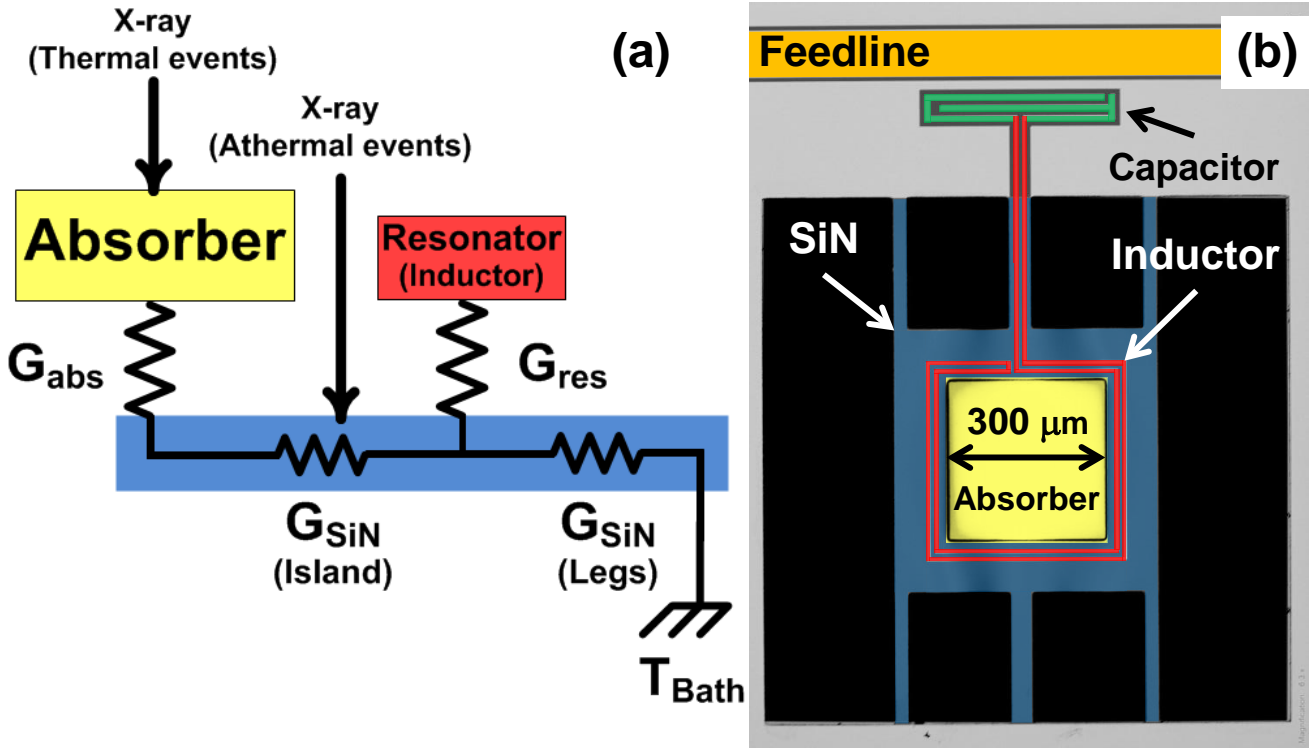

FIG. 1. (Color online) (a) A thermal circuit diagram of the tantalum device showing the dominant thermal conductances and locations for X-ray absorption. The electron-phonon couplings and electronic heat conductance through the inductor are not shown. (b) A false-color micrograph of the tantalum device. The lumped element resonator is fabricated lithographically from a 100 nm thick $WSi_2$ film (gray). The interdigitated capacitor (IDC) portion (green) is on solid SiN/Si substrate (500 nm / 300 μm), while the inductive meander (red) lies on a suspended SiN membrane (blue). The meander encircles the absorber (yellow). The coplanar waveguide (CPW) feedline is shown in yellow and the ground plane in gray.

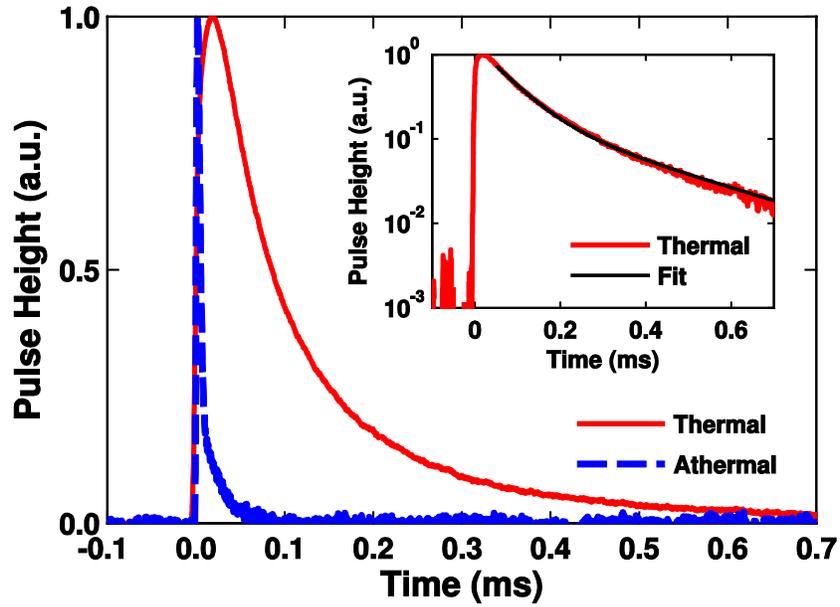

FIG. 2. (Color online) Pulse templates obtained by averaging 20 pulses from 6 keV photons. The blue dashed curve represents the typical fast response for photon absorption directly in the resonator or in the SiN membrane. The red curve represents the thermal response for photons absorbed in the tantalum absorber. The insert shows the thermal template fit with two exponentials.

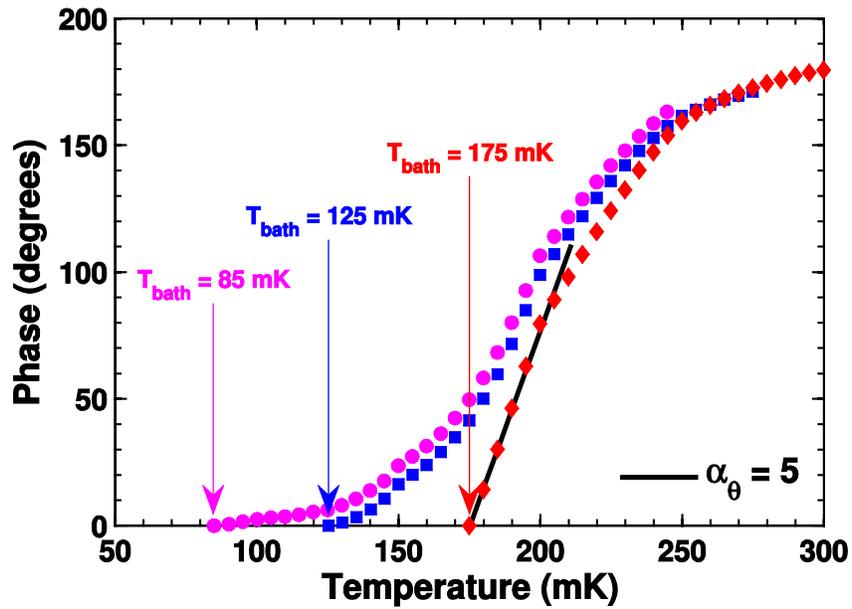

FIG. 3. (Color online) Phase variation induced in the resonator by sweeping the cryostat temperature, starting at different $T_{bath}$ (85 mK magenta circles, 125 mK blue squares, and 175 mK red diamonds). These data represent the calibration curves of the resonator as a thermometer. The black line is the logarithmic sensitivity, $\alpha_\theta$ at $T_{bath}$=175 mK.

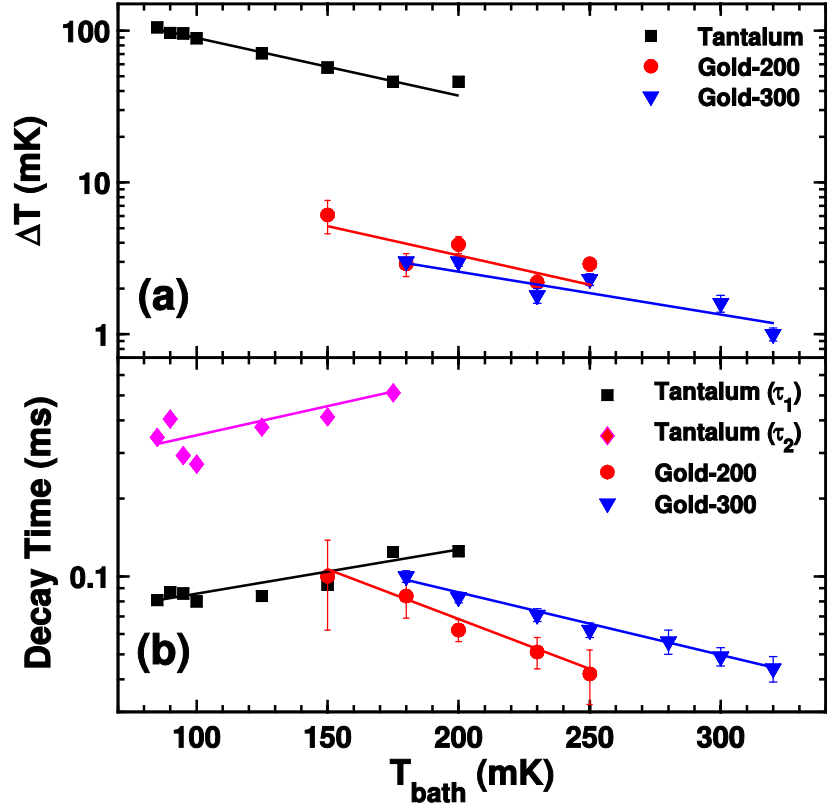

FIG. 4. (Color online) (a) Temperature rise (ΔT) induced in the resonator by the absorption of a 6 keV photon in a square gold absorber of 200 μm side (red circles), 300 μm side (blue triangle) and a square tantalum absorber of 300 μm side (black square) at different $T_{bath}$. (b) Decay times of the thermal pulses at different $T_{bath}$ for both gold and tantalum absorbers. Red circles represent the 200 μm gold absorber, while blue triangles represent the 300 μm gold absorber one. For the tantalum absorber, the two decay times present in the thermal pulses are shown for a single size absorber (300 μm side). The solid lines are guides for the eye.